# Field-effect-induced two-dimensional electron gas utilizing modulation doping for improved ohmic contacts

Sumit Mondal[1,2], Geoffrey C. Gardner[1,3], John D. Watson[1,2], Saeed Fallahi[1,2] Amir Yacoby[5], and Michael J. Manfra[1,2,3,4,*]

[1] Birck Nanotechnology Center, Purdue University, West Lafayette IN, 47907

[2] Department of Physics and Astronomy, Purdue University, West Lafayette IN, 47907

[3] School of Materials Engineering, Purdue University, West Lafayette IN, 47907

[4] School of Electrical and Computer Engineering, Purdue University, West Lafayette IN, 47907

[5] Department of Physics, Harvard University, Cambridge MA, 02138

**Modulation-doped AlGaAs/GaAs heterostructures are utilized extensively in the study of quantum transport in nanostructures, but charge fluctuations associated with remote ionized dopants often produce deleterious effects. Electric field-induced carrier systems offer an attractive alternative if certain challenges can be overcome. We demonstrate a field-effect transistor in which the active channel is locally devoid of modulation-doping, but silicon dopant atoms are retained in the ohmic contact region to facilitate reliable low-resistance contacts. A high quality two-dimensional electron gas is induced by a field-effect and is tunable over a wide range of density. Device design, fabrication, and low temperature (T= 0.3K) transport data are reported.**

Nanostructures such as quantum point contacts and quantum dots fabricated on modulation-doped AlGaAs/GaAs heterostructures are widely used to explore nanoscale electron transport and are utilized extensively in spin-based approaches to quantum computing [1-8]. Modulation-doped GaAs/AlGaAs heterostructures possess several desirable attributes including very high mobility of the underlying two-dimensional electron gas (2DEG) and the relative ease of nanostructure fabrication. However, the presence of ionized dopants inherent to modulation-doping can have adverse effects on the behavior of nanostructures and are a possible source of decoherence for spin-qubits [9, 10]. Ionized dopants can act as active trapping sites for electrons injected from the metal surface gate through the Schottky barrier [11], giving rise to random switching of the charge state of the impurities. These fluctuations cause instability through a time-dependent potential landscape [10-13]. Methods such as 'bias cooling' in which nanostructures are cooled while a positive bias applied to the gates aid in reducing fluctuations [11], but charge noise is still believed to a dominant mechanism limiting gate fidelities in spin qubits [9].

Field-effect-induced 2DEG transistors (FETs) in GaAs/AlGaAs heterostructures have been investigated extensively [14-27] and might find utility as a platform to investigate nanoscale phenomena in a low-noise environment if certain limitations can be overcome. The most widely studied device is the heterostructure-insulated-gate field-effect transistor (HIGFET), in which a highly-conducting n+ GaAs gate is grown on top of an insulating $Al_xGa_{1-x}As$ barrier layer by molecular beam epitaxy (MBE) [14-16, 18, 22]. HIGFET fabrication requires placing ohmic contacts in intimate contact with the primary AlGaAs/GaAs interface where the 2DEG will reside without shorting the ohmic contact to the n+ GaAs gate [17, 18]. This challenging task becomes increasingly difficult for shallow 2DEG structures required for many nanostructure devices [15]. Device yield is often low and the devices can suffer from high contact resistance. Another FET design reported in the literature utilizes a lithographically-defined global top-gate deposited on top of insulators such as polyimides [17, 19-20, 23], nitrides [24, 25] or oxides [26] separating the ohmic contacts from the gate. In this configuration the global top-gate must extend over the ohmic contact pad to ensure a continuous 2DEG to the ohmic contact. Progress has been made in creating a 2DEG on shallow undoped structures by this method [20, 21] and the induced 2DEG can have high mobility [24, 25] in deeper structures. Nevertheless, the requirement that the gate electrode overlaps the rough annealed ohmic metal enhances the possibility of undesired leakage paths [27, 28]. In general, the absence of dopants makes fabrication of reliable ohmic contacts with high yield and low contact resistance challenging [20, 28] in completely undoped structures.

In this letter we demonstrate an alternative FET device design that is simple to fabricate, has reliable ohmic contacts and negligible gate leakage. It is devoid of the dopant atoms in the vicinity of the active channel but retains silicon doping in the region of ohmic contacts yielding low contact resistance. An AlAs etch-stop layer designed into the epilayer facilitates removal of the modulation-doping layer in specific locations by selective wet etching. The density of the field-induced 2DEG can be modulated from $6.5 \times 10^{10} cm^{-2}$ to $2.7 \times 10^{11} cm^{-2}$ and the 2DEG exhibits peak mobility of $2.4 \times 10^6 cm^2/Vs$ at T=0.3K. The observation of extremely low hysteresis of the 2DEG density and mobility during multiple gate voltage sweeps indicates that this device design will yield stable nanostructures.

The underlying heterostructure of this device is grown by MBE. The 2DEG resides at a single $GaAs/Al_{0.26}Ga_{0.74}As$ heterojunction. Fig. 1(a) and (b) respectively, represent the schematic cross-section and the simulated bandstructure for the as-grown wafer. The heterostructure consists of a 1µm thick GaAs channel grown on a (100) GaAs substrate followed by a 175nm thick $Al_{0.26}Ga_{0.74}As$ barrier and a 10nm GaAs cap. It is delta-doped with silicon to $1.85 \times 10^{12}$ cm$^{-2}$ at a setback of 75nm above the primary $GaAs/Al_{0.26}Ga_{0.74}As$ interface. The salient feature of this structure is the presence of a 3nm AlAs etch stop layer situated 10nm below the doping layer. In the as-grown wafer the modulation-doped silicon layer results in the formation of a uniform 2DEG as shown in Fig. 1(b). However, the presence of the AlAs etch stop allows for the removal of the silicon doping layer, and hence the 2DEG, in specific sites defined by lithography. For the devices discussed here, a simple Hall bar geometry is employed.

Device fabrication proceeded as follows. 10 small mesas were defined with standard photolithography. 5 of these mesas are shown in Fig. 1(d) which represents one-half of a full device. These mesas serve as the ohmic contacts and access connections to the field-induced 2DEG. After a dip in dilute buffered-oxide-etch (BOE) to desorb the surface oxide, the patterned chip was etched using a 6:1 citric acid: $H_2O_2$ mixture. The entire chip was etched past the Si doping layer to the

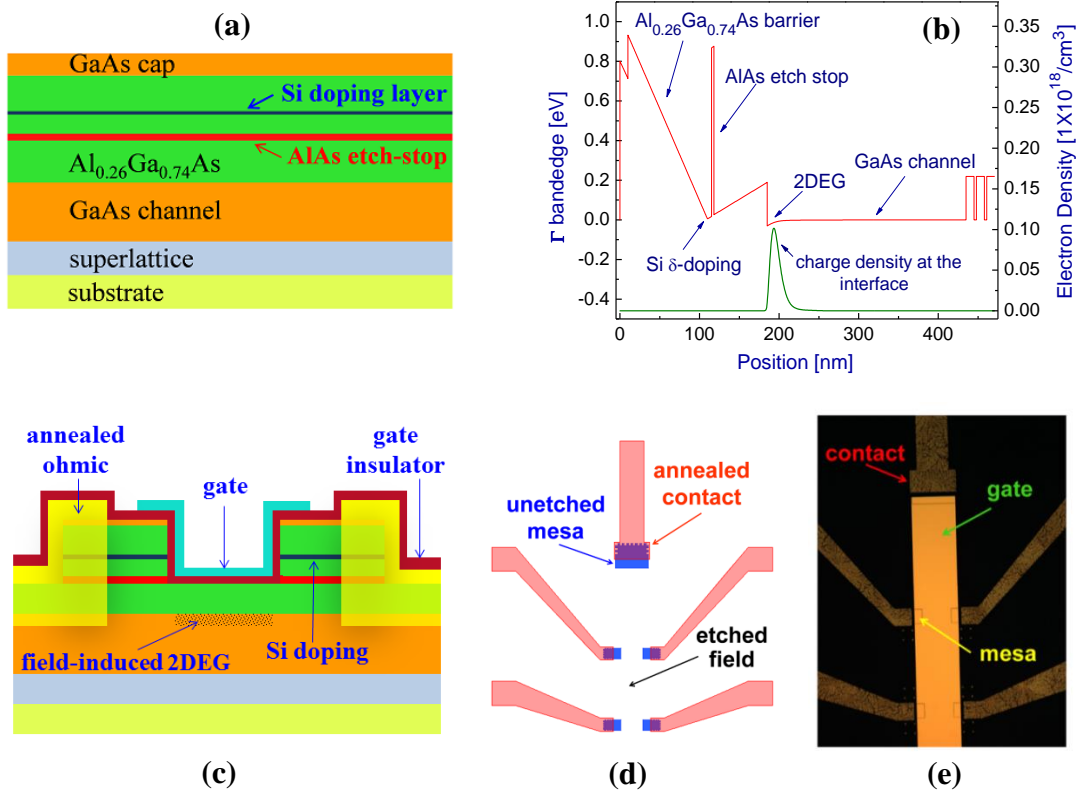

Figure 1 (a) Schematic cross-section of the MBE-grown wafer (b) Simulated bandstructure of the as-grown heterostructure using Nextnano3 [33] (c) Schematic cross-section of the device design (d) Schematic top-view of a portion the FET, the top-gate is not shown for clarity (e) Optical micrograph of the fabricated FET

AlAs etch stop, except for the 10 small mesas. The etch rate of the citric acid: $H_2O_2$ mixture was extremely sensitive to x, the mole fraction of $Al_xGa_{1-x}As$ [29]. The citric acid: $H_2O_2$ mixture etched the x=0.26 barrier at ~4.5nm/sec but stopped at the pure AlAs etch-stop immediately below dopant layer. Dilute BOE, which is highly selective in etching AlGaAs with mole fraction above x=0.4, was used to remove the sacrificial 3nm AlAs layer. In the etched field that forms the conducting channel of our device, only the undoped 62nm $Al_{0.26}Ga_{0.74}As$ barrier remains above the GaAs buffer. The contact mesas and the etched field were carefully investigated using atomic force microscopy for any defects or non-uniformity of surface morphology. Sub-nanometer surface roughness in the etched field was achieved. Ni/Ge/Au (30nm/60nm/120nm) contacts were evaporated on part of the mesa regions and subsequently annealed at $440^0C$ for 2 minutes in a forming gas atmosphere to diffuse the alloy into the 2DEG. A 63nm silicon nitride ($Si_3N_4$) gate insulator was then deposited at 300°C utilizing a multistep deposition process to prohibit the upward propagation of pinhole defects. The final fabrication step was the deposition of the Ti/Au (15nm/160nm) gate over the insulating $Si_3N_4$ layer. Fig. 1(c) and (d) show a schematic cross-section of the device design and the schematic top-view of the Hall-bar FET. It is important to note that the gate does not overlap the ohmic metal layer. This can be clearly seen in the optical micrograph of our completed device (Fig. 1(e)). This separation eliminates a potential leakage path between the gate and the annealed ohmic contact. The small region between the annealed ohmic and the active field-effected 2DEG retains its modulation-doping, however, providing a low resistance path.

The completed FET was cooled to low temperature to characterize device operation, contact resistance, and magnetotransport quality. All of the measurements reported here were performed without illumination at the base temperature (T=0.3K) of a Janis top-loading $He^3$ system using standard AC lock-in techniques at a frequency of 13Hz with a 500nA excitation current. Consistent with the operational principle of an enhancement-mode FET, no conduction was observed in the device until the gate was energized above a threshold voltage. The onset of conduction was observed at a gate bias of 0.95V, roughly consistent with the potential required to overcome Fermi-level pinning at the semiconductor/$Si_3N_4$ interface. The GaAs/AlGaAs heterojunction was populated with an increasing density of carriers coming from the ohmic contacts as the gate was biased gradually beyond this threshold. The density was determined from the position of integer quantum Hall minima in the longitudinal resistance ($R_{xx}$). The density increased linearly with increasing gate bias and could be tuned from ~$6.5\times10^{10}$ cm$^{-2}$ to ~$2.6\times10^{11}$ cm$^{-2}$ with capacitance per area $3\times10^{11}$ cm$^{-2}$/V at T=0.3K (Fig. 2(a)). This behavior indicated strong capacitive control of the 2DEG. The gate leakage was at or below 15pA for all applied biases, the measurement noise-floor of this experiment. The variation of the mobility of the FET with the 2DEG density is shown in Fig. 2(b). It rose from $4\times10^5$ cm$^2$/Vs at the onset of conduction to $2.4\times10^6$ cm$^2$/Vs, monotonically increasing with density. The highest mobility was comparable to the measured mobility of the underlying modulation-doped structure prior to any processing ($2.6\times10^6$ cm$^2$/Vs at T=0.3K) indicating minimal processing related degradations. Insight about the dominant scattering mechanisms occurring in the channel can be obtained through examination of the density dependence of the mobility. The mobility ($\mu$) was found to vary with density (n) as $\mu \propto n^{\alpha}$ with the exponent $\alpha=1.03\pm0.02$ indicating the dominance of unintentional background impurity scattering in this structure [30, 34]. Figure 3(a) displays transport in the quantum Hall regime at a gate voltage of 1.1V. The high quality of the field-induced 2DEG is manifested in vanishing longitudinal resistance $R_{xx}$ concomitant with plateaus in $R_{xy}$, corresponding to fully developed integer quantum Hall states. This data also confirmed the absence of any parallel conducting channel in the device. A more thorough investigation of the transport properties and noise characteristics of these devices will be presented in a future publication.

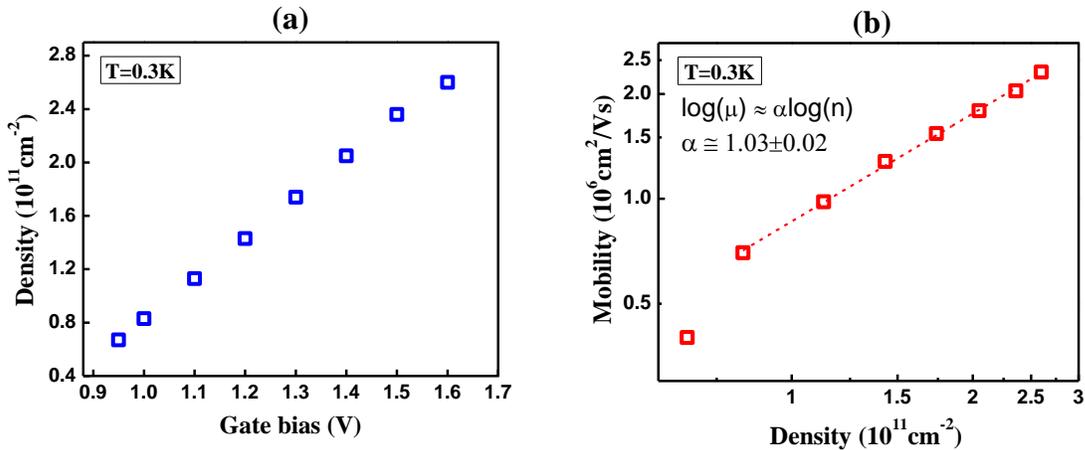

Figure 2(a) Gate voltage dependence of the 2DEG density of the FET (b) Density dependence of the 2DEG mobility of the FET on a log-log scale

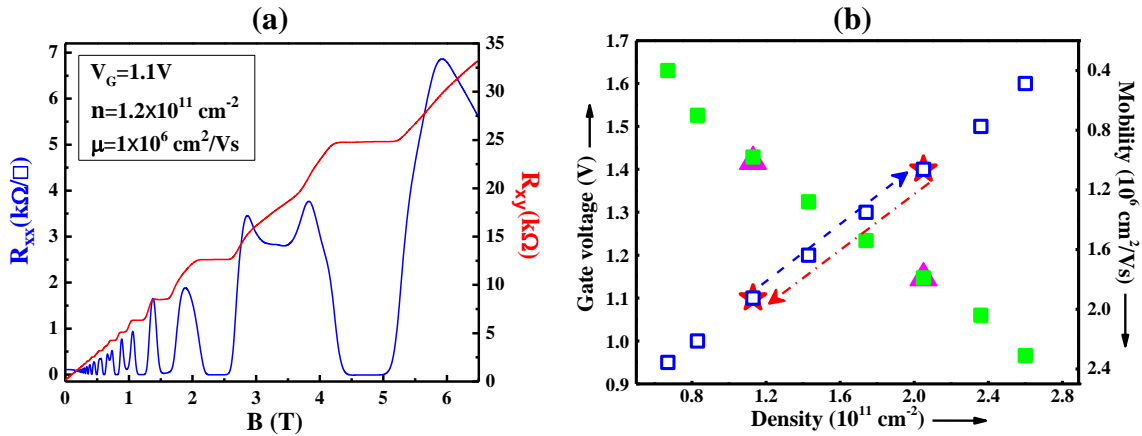

Figure 3(a) Magnetotransport of the FET at a density $1.2\times10^{11}$ cm$^{-2}$ (b) Stability of the device under gate operation. The open blue (closed green) squares represent the density (mobility) while the gate was swept to higher bias. The red stars (pink triangles) represent the density (mobility) as the gate was swept to lower bias.

Reproducibility of the device is documented in Figure 3(b). The 2DEG density was increased by ramping the gate from 0.95 Volt to 1.65 Volts (marked by the open blue squares in Fig. 3(b)) and then the gate potential was decreased which resulted in a non-hysteretic, highly repeatable, reduction in density (indicated by representative closed red stars in Fig. 3(b)). Similar repeatability was observed for the mobility (marked by the green squares and pink stars in Fig. 3(b)) in this range. This behavior indicates minimal gate-induced rearrangement of charge at the semiconductor-insulator interface. In addition, the low temperature electrical properties of the device were found to be completely reproducible over the course of several thermal cycles to room temperature. The contact resistance of the device was measured using a three-terminal method described in Ref. [32]. The measured contact resistance was ~70Ω. Since the gate does not overlap the ohmic metal in the device, the measured contact resistance is a sum of contributions from the annealed metal contact as well as the portion of mesa that extends from the ohmic metal to the gated region of the device. We note that low contact resistance ($R_c \leq 100\Omega$) is typically a necessary condition for high speed readout of spin qubits, as parasitic RC time constants can limit measurement bandwidth [35].

In conclusion, we have experimentally demonstrated the fabrication and operation of a FET device that utilizes modulation doping for reliable ohmic contact formation, but removes the dopant atoms in the active channel. The fabrication is relatively simple and high-yield. The resulting 2DEG has high mobility and is widely tunable in density from ~$6.5\times10^{10}$ cm$^{-2}$ to $2.6\times10^{11}$ cm$^{-2}$ with peak mobility $2.4\times10^{6}$ cm$^{2}$/Vs. The device is non-hysteretic, reproducible upon thermal recycling, and has minimal gate leakage. While a simple Hall bar geometry was explored here, this architecture is amenable to nanostructure formation with additional depleting surface gates. Fabrication of nanostructures and study their noise properties are currently underway.

*email: mmanfra@purdue.edu

**Acknowledgements**

This work was supported by the Office of the Director of National Intelligence, Intelligence Advanced Research Projects Activity (IARPA), through the Army Research Office grant W911NF-12-1-0354. Y.A. is also supported by the United States Department of Defense. The views and conclusions contained in this document are those of the authors and should not be interpreted as representing the official policies, either expressly or implied, of the U.S. Government.